\documentclass[twocolumn,showpacs,preprintnumbers,amsmath,amssymb]{revtex4}

\usepackage{graphicx}
\usepackage{dcolumn}
\usepackage{bm}

\begin{document}


\title{Nuclear Magnetic Orderings and Frustration in Bcc $^3$He in High
Magnetic Fields}

\author{Hiroshi Fukuyama$^1$$^{\ast}$, K. Yawata$^2$, 
T. Momoi$^2$$^{\dagger}$, H. Ikegami$^3$$^{\dagger}$, and H. Ishimoto$^3$}

\affiliation{$^{1}$Department of Physics, Graduate School of Science, University of Tokyo, 7-3-1 Hongo, Bunkyo-ku, Tokyo 113-0033, Japan\\
$^{2}$Institute of Physics, University of Tsukuba, 1-1-1 Ten-nodai, Tsukuba, Ibaraki 305-0006, Japan\\
$^{3}$Institute for Solid State Physics, University of Tokyo, 5-1-5 Kashiwanoha, Kashiwa-shi, Chiba 277-8581, Japan
}%


\date{\today}

\begin{abstract}
We studied phase transitions and thermodynamic properties of the 
field-induced antiferromagnetic nuclear-spin ordered phase in bcc solid 
$^3$He by measuring the melting pressure at temperatures down to 0.4 mK in 
magnetic fields up to 15 T.
The transition temperature from the paramagnetic phase is 
reentrant with increasing field with a maximum at 10 T.
This indicates that the system is highly 
frustrated by the competing multiple-spin exchanges and 
excludes other mechanisms related to the zero-point vacancies.
The upper critical field was estimated as 19.7 T,
which reveals non-negligible contributions from
higher order exchanges beyond six spins.
A considerable softening of the spin waves in the ordered phase 
in low fields also suggests the strong frustration.
\end{abstract}

\maketitle

Solid $^3$He with bcc structure is an ideal three dimensional antiferromagnet
with intrinsic frustration due to the competing multiple-spin exchanges 
(MSEs) \cite{Roger1983}.
Nearly perfect bcc crystals can easily be grown when they coexist with the 
liquid phase.
It has a two dimensional (2D) counterpart, monolayer solid 
$^3$He adsorbed on graphite, where an exotic ground state so called 
the \emph{gapless quantum spin liquid} is realized \cite{Ishida}.
Thus we can study the frustration due to the MSEs and its dimensionality effects 
in great details through investigations of nuclear magnetism of 
bcc and 2D $^3$He.

The exchange interactions among the nuclear spins ($S = 1/2$) in solid 
$^3$He are associated with the purely isotropic direct atom-atom exchanges.
They completely dominate the thermodynamic properties below 100 mK, 
which makes this material an exceptionally pure magnetic system.
As Thouless \cite{Thouless} first conjectured, the frustration in this 
system originates from the competition between antiferromagnetic (AFM) 
and ferromagnetic (FM) MSEs.
He showed that even numbers of exchanging atoms introduce AFM interactions 
while odd numbers do FM ones.
The importance of the four-spin ring exchange has recently been recognized 
in highly correlated electronic systems as well \cite{ring-exchange}.

It is a remarkable and unique feature of solid $^3$He that even 
absolute values of the MSE frequencies ($J_P$) are potentially 
calculable from the first principles.
$J_P$ up to six-spin exchanges have been eventually calculated by the path 
integral Monte Carlo (PIMC) technique \cite{Ceperley} for bcc $^3$He 
at a molar volume (24.12 cm$^3$/mol) close to the melting one.
They are consistent with most of the so far existing experiments 
semi-quantitatively (within 30\%) \cite{Fukuyama-Bc2,Fukuyama-Bc1}.
It is, however, generally difficult to test them more rigorously because of the 
many parameters involved.
In addition, theoretical approximations applied to the MSE Hamiltonian 
sometimes limit the accuracy of the tests.
Therefore, there still remains a fundamental question for the MSE model; 
\emph{Where can we truncate the series of the MSE interactions?}

The competing interactions in bcc $^3$He might be explained by quite 
different hypotheses from the MSE model.
The most intriguing one is so called the \emph{zero-point vacancy 
(ZPV) model} \cite{Andreev}.
If the system contains a finite amount of ZPVs, 
FM interactions will be induced among nuclear spins surrounding 
the ZPVs in order to reduce their kinetic energies.
Such FM interactions would compete with the ordinary AFM two-spin 
exchange interaction.
This model can be quantitatively tested by exploring the high-field magnetic 
phase diagram at the melting density where the ZPV concentration is maximized.
This test is of up-to-date interest since ZPV could be responsible for 
the superfluid responses recently observed in solid $^4$He \cite{supersolid}.
It is also important in the light of determining
the zero-temperature upper critical field $B_{c2}$(0), beyond which 
the spins are fully polarized along the external field, for this field 
is exactly calculable even with the mean field theory \cite{Fukuyama-Bc2}.

In this Letter, we present results of melting pressure measurements of 
bcc $^3$He down to temperatures well below the AFM ordering 
temperature ($T_c$) and in high magnetic fields up to 15 T, a factor of two 
higher field than in the previous experiment \cite{Godfrin1980-MC}.
The melting pressure ($P$) is a useful probe to study thermodynamic properties
and phase transitions in $^3$He.
The temperature variation of $P$ is determined by the Clausius-Clapeyron
relation:
\begin{eqnarray}
dP / dT= (S_l - S) / (V_ l - V),
\label{eq.CC}
\end{eqnarray}
where $S$ ($S_l$) and $V$ ($V_l$) are the molar entropy and volume of the 
solid (liquid) phase, respectively.
Note that, below 5 mK, $S_l$ is less than 10\% of $S$ and precisely known
from the specific heat measurement \cite{Greywall-T-scale}.
The volume difference $(V_{l} - V) =$ 1.314 cm$^3$/mol \cite{Halperin} is 
expected to be temperature and field independent.
Thus the melting pressure change is determined predominantly by the entropy
of the solid phase.
Details of the experimental setup have already been described elsewhere 
\cite{Yawata}.
We took all the data under complete thermal equilibrium.
\begin{figure}
\includegraphics[width = 6.5 cm]{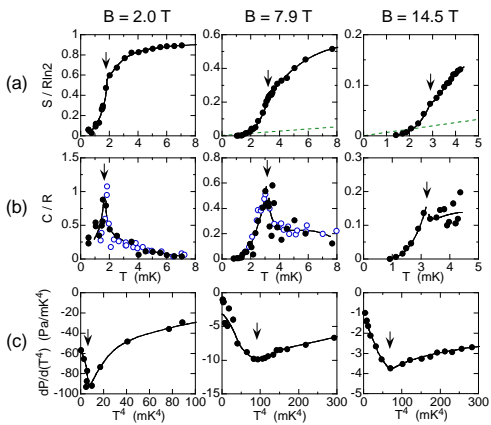}
\caption{(a) Entropy, (b) specific heat and (c) $\partial P /
\partial (T^{4})$ of bcc $^3$He deduced from the melting pressure ($P$) 
data at $B = 2.0$, 7.9 and 14.5 T.
The vertical arrows indicate the HFP-PP transitions.
The solid lines are guides to the eye.
The open circles are from the previous direct specific-heat measurements 
\cite{Sawada}.}
\label{fig.transition}
\end{figure}

Figure 1 (a)(b) show the entropy and the specific heat ($C$) of bcc $^3$He 
deduced from the measured melting pressure through Eq.\ (\ref{eq.CC}) at 
three selected fields. The HFP-PP transitions are identified as kinks in 
the $S$ vs.\ $T$ plots and as peaks in the $C$ vs.\ $T$ plots.
These features indicate that the transitions are not first order but 
continuous.
This is consistent with the direct specific heat measurements below 8 T
\cite{Sawada,Greywall1987}, and with the theoretical proposal that 
the spin structure of HFP is canted normal antiferromagnetic (CNAF) 
with two sublattices \cite{Roger1983} and cubic symmetry \cite{Osheroff-HFP} 
(see Fig.\ 2).
$T_c$ can also be determined as a temperature where $\partial P / \partial 
(T^{4})$ has a minimum in the plot as a function of $T^4$ (Fig.\ 1c).
The $T^4$ dependence of $P(T)$ is expected from the AFM spin wave theory in 
the ordered phase.
The above three different determinations of $T_c$ agree each other 
within 5\% except for 2 and 6 T (13 and 16\%, respectively).

Figure 2 shows the magnetic phase diagram of bcc $^3$He at 24.21 
cm$^3$/mol. The data points denote phase boundaries between two of the 
three magnetic phases (HFP, LFP and PP) determined by the present and 
previous experiments
\cite{Fukuyama-Bc2,Sawada,Osheroff-HFP,unpublished,Godfrin1980-MC,Xia}.
Here the low-field phase (LFP) is another AFM ordered phase with four 
sublattices (the U2D2 phase \cite{OCF}) which exists below the lower 
critical field ($B_{c1} = 0.45$ T).
Note that we 
made small volume corrections for the original data so as to be 
consistent with the result for the fixed $V$ at 24.21~cm$^3$/mol
in Figs. 2, 3(b) and 4.
This was necessary because $V$ and hence $J_P$ increase slightly 
with increasing $B$ along the melting curve.
The largest correction is 4.6\% at $B = 14.5$~T.
\begin{figure}
\includegraphics[width = 8.5 cm]{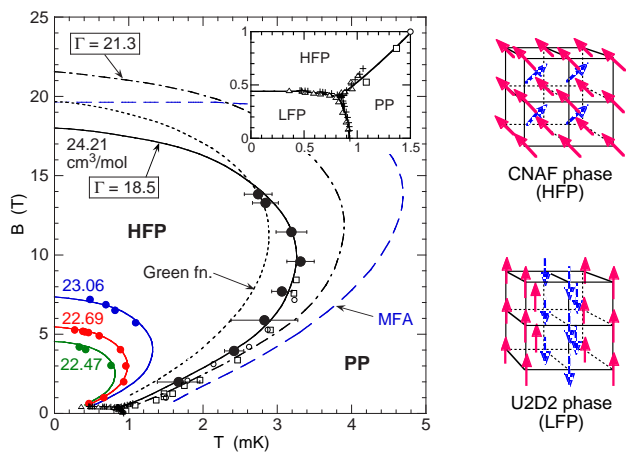}
\caption{Nuclear magnetic phase diagram of bcc $^3$He at 24.21 cm$^3$/mol.
The data points are phase transitions between two of the three phases (HPF, 
LFP and PP); this work ($\bullet$), Ref.\ \cite{Godfrin1980-MC} ($\circ$),
Ref.\ \cite{Sawada} ($\square$), Ref.\ \cite{Osheroff-HFP} ($\triangle$)
and Ref.\ \cite{Xia} ($+$).
Also shown are the HFP-PP transitions at 22.47, 22.69 and 
23.06 cm$^3$/mol from Ref.\ \cite{Fukuyama-Bc2,unpublished}.
The solid and dash-dotted lines are HFP-PP boundaries scaled from that 
determined for 22.69 cm$^3$/mol with $\Gamma =$ 18.5 and 21.3, respectively.
For the other lines see the text.}
\label{fig.PD}
\end{figure}

Obviously the HFP-PP boundary has a reentrant shape with a maximum 
$T_c$ ($=3.4$~mK) near $B= 10$~T.
This is the first experimental evidence that $T_c$ turns around to decrease 
above a certain field at the melting density as was observed at 
a much higher density \cite{Fukuyama-Bc2}.
From this, we conclude that the HFP is an AFM ordered phase 
and not any phase with high polarization associated 
with the ZPV.
Within the ZPV model \cite{Andreev}, upon increasing magnetic field, 
the vacancy band width increases due to the increasing polarization and the 
vacancy creation energy decreases due to the decreasing melting pressure.
Hence the vacancy concentration $n$ should monotonously increase, which 
never yields the negative slope for the phase boundary.
Generally, the Gr\"{u}neisen constant for $n$ should be quite different 
from that for $J_P$, i.e., $\Gamma (J_P) \equiv \partial \ln J_P/ \partial 
\ln V \approx 18$ \cite{Fukuyama-Bc1}.
Nevertheless, the measured HFP-PP boundary has a similar volume 
dependence to $\Gamma (J_P)$ as the data points follow approximately the 
solid line estimated from the boundary at 22.69 cm$^3$/mol \cite{Fukuyama-Bc2} with $\Gamma =$ 18.5 (see Fig. 2).

The dotted line in Fig.\ 2 is a CNAF-PP boundary calculated by the Green 
function method \cite{Iwahashi} with the following MSE parameters:
\begin{eqnarray}
J_{1N}& = &0.48, J_{2N} = 0.067, T_{1} = 0.20, K_{P} = 0.28, \nonumber \\
K_{F}& = &0.028, K_{A} = 0.006, K_{B} = 0.0005, K_{L} = 0.011, \nonumber \\
K_{S}& = &0.0020, S_{1} = 0.037, S_{2} = 0.023,
\label{eq.J}
\end{eqnarray}
given by the PIMC calculation \cite{Ceperley}.
Here, all numbers are in mK, and we followed Ref.~\cite{Godfrin1988-MSE}
for the notations of $J_P$; 
two-spin ($J_{1N}$, $J_{2N}$), three-spin 
($T_{1}$), four-spin ($K_{P}$, $K_{F}$, $K_{A}$, $K_{B}$, $K_{L}$, 
$K_{S}$) and six-spin ($S_{1}$, $S_{2}$) exchanges.
Note also that all the original MSE parameters in Ref.\ \cite{Ceperley} have 
been increased only by 4\% in Eq.\ (\ref{eq.J}) so that they give 
$B_{c2}(0) =$ 19.7 T, the estimation discussed later.
The agreement between the Green function calculation and the present data 
is good.
The dashed line is the HFP-PP boundary calculated by the mean 
field approximation (MFA).
It gives too high $T_c$ compared to the experiment particularly at 
high fields indicating the importance of thermal fluctuations \cite{Fukuyama-Bc2}.
\begin{figure}
\includegraphics[width = 8.8 cm]{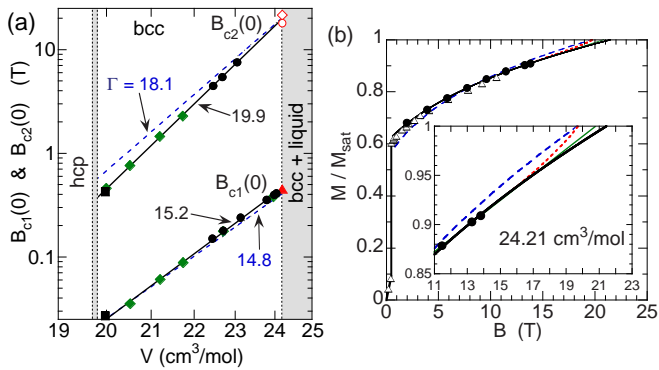}
\caption{(a) Log-log plots for the zero-temperature upper and lower critical fields, 
$B_{c2}$(0) and $B_{c1}$(0), as functions of $V$; 
Ref.\ \cite{Fukuyama-Bc2,unpublished} ($\bullet$), 
Ref.\ \cite{Xia,Osheroff1987-MCinHF} ($\blacktriangle$), 
Ref.\ \cite{Okamoto} ($\blacksquare$), 
Ref.\ \cite{Omelaenko} ($\blacklozenge$), 
estimations from the HFP-PP boundary ($\circ$) and 
Eq.\ (\ref{eq.magcurv}) ($\lozenge$).
The dashed lines are MFA calculations with the MSE parameters
of Eq.\ (\ref{eq.J}) and their volume dependencies given in Ref.\ \cite{WKB,RH}.
(b) Magnetization of bcc $^3$He at $T =$ 0; this work 
($\bullet$) and Ref.\ \cite{Osheroff1987-MCinHF} ($\triangle$).
The solid line represents Eq.\ (\ref{eq.magcurv}) and
the thin line is the linear extrapolation of the highest three data points.
The dashed line is the MFA calculation with Eq.\ (\ref{eq.J}).
The dotted line is a guide to the eye giving $B_{c2}(0) =$ 19.7 T.}
\label{fig.magcurv}
\end{figure}

An accurate estimation for $B_{c2}$(0) at the melting density 
is available by extrapolating the volume dependence of this quantity 
which is known more reliably at higher densities.
As is shown in Fig. 3(a), all the existing data 
\cite{Fukuyama-Bc2,unpublished,Okamoto,Omelaenko} 
follow the power law dependence with $\Gamma = 19.9 \pm 0.2$ 
very well, which gives $19.7 \pm 0.4$ T at 24.21 cm$^3$/mol.
This Gr\"{u}neisen constant is considerably larger than those for 
any other physical quantities measured at lower fields ($\approx 18$).
The WKB calculations \cite{WKB,RH} predict the larger $\Gamma$ for 
the smaller $J_P$ in the density range for bcc $^3$He,
i.e., $\Gamma (J_{1N}) =$ 18.2, $\Gamma (T_1) =$ 16.2,
$\Gamma (K_P) =$ 15.1, $\Gamma (K_F) =$ 18.1, and $\Gamma (S_1) =$ 18.9 
which explain very well the exceptionally small experimental 
$\Gamma$ value ($= 15.2 \pm 0.2$) for $B_{c1} (0)$ 
\cite{Fukuyama-Bc1,Omelaenko}.
On the other hand, the same WKB calculations predict $\Gamma =$ 18.1
for $B_{c2} (0)$ which is too small compared to the experiment (see Fig. 3(a)).
It is important to recognize that higher order exchanges contribute 
more effectively to magnetic properties at higher fields 
particularly to $B_{c2}(0)$~\cite{Ceperley,Godfrin1988-MSE}.
The discrepancy can not be explained even by infinitely large $\Gamma$
values for $S_{1}$ and $S_{2}$ in Eq.\ (\ref{eq.J}).
We thus conclude that the convergence of the series of MSE interactions
is somewhat slower than the current assumption and that 
higher order exchanges than six spins should be taken into account.
If so, at fields above 15 T, the $B_{c2}(T)$ line would deviate upward from 
the solid line in Fig. 2, which assumes the constant $\Gamma$ value (= 18.5),
in order to smoothly connect to 19.7 T at $T =$ 0.
Note that the ZPV model gives a lower $\Gamma$ value than 18 because of
the increasing FM contribution at larger volumes, which is inconsistent
with the present result.

It is intriguing to note that $T_c$ approaches zero as $B\rightarrow 0$
(inset of Fig. 2). 
In other words, if the LFP did not exist hypothetically, 
the system would have no long range order at $B =$ 0 down to zero 
temperature due to the strong frustration.
Actually, this is realized in 2D $^3$He \cite{Ishida}.
Thus a similar high-field ordered phase may exist in 2D as well \cite{2D}.
Another point is that even the highest $T_c$ ($= 3.4$ mK) is five times 
lower than the exchange energy represented by $\mu B_{c2}(0)$ ($= 16$ mK), 
where $\mu$ is the nuclear magnetic moment of $^3$He.
This large suppression of $T_c$ is a further evidence for the strong 
frustration inherent in this system.

We extrapolated the melting pressure data below 0.6 $T_c$ to $T=0$ by 
fitting them to
\begin{eqnarray}
P (T, B) = P (0, B) - cT^4.
\label{eq.SW}
\end{eqnarray}
The zero temperature magnetization $M$ was deduced from the 
fitted $P (0, B)$ using the magnetic Clausius-Clapeyron eq. as was 
done in Ref.\cite{Osheroff1987-MCinHF}.
The reduced magnetization, $m \equiv M / M_{sat}$, obtained in this way
is shown in Fig. 3(b). 
Here $M_{sat}$ is the saturation magnetization. 
Our data can be fitted well to the thick solid line expressed by
\begin{eqnarray}
B = -(10.1 \pm 1.7)m + (25.4 \pm 5.2)m^3 + (6.2 \pm 3.9)m^5
\label{eq.magcurv}
\end{eqnarray}
with an rms deviation of 1.5\%. 
This functional form is an MFA expression for the magnetization curve 
based on the MSE Hamiltonian considering up to the six-spin exchanges 
\cite{Godfrin1988-MSE}.
A simple extrapolation of Eq.\ (\ref{eq.magcurv}) to $m$ = 1 gives
$B_{c2}(0) = 21.5 \pm 0.4$ T, which agrees well with the similar extrapolation
in Ref.\ \cite{Osheroff1987-MCinHF} ($= 21.7 \pm 1$ T) but is
higher than the above mentioned estimation (= 19.7 T) by 9\%.
This is an additional indirect evidence for the contributions from 
the currently ignored higher order exchanges.
The magnetization seems to have a weak positive curvature
above 15 T in order to be consistent with $B_{c2}(0) =$ 19.7 T \cite{fluctuations}.
One should be careful about the accuracy of MFA, which ignores 
fluctuations, even near $B_{c2}$(0) since it gives too high $B_{c2}$
at finite temperatures.
However, a better approximation taking account of fluctuations such as 
the coupled cluster approximation (CCA) gives essentially a similar 
magnetization curve to the MFA one \cite{RH}. 

From the fitting parameter $c$ in Eq.\ (\ref{eq.SW}), we deduced the 
angle averaged spin-wave velocity ($v$) using the relation:
\begin{eqnarray}
c = \frac{N\pi^2 k_B^4}{90 \hbar^3} \frac{V}{(V_l - V)}\frac{1}{v^3}.
\label{eq.SWvelocity}
\end{eqnarray}
The number of spin-wave modes ($N$) is assumed as unity, since thermal 
energies at $T \leq 0.6 T_c$ are much lower than the Zeeman 
energies in the field range studied here.
In Fig.\ 4 we plot the deduced $v$ as a function of $B$. 
The field dependence is qualitatively similar to that of $T_{c}(B)$, because $T_c$ 
is a measure of the spin-wave stiffness.
The smooth extrapolation to lower fields agrees well with 
the previous data below 1 T \cite{Greywall1987,Osheroff1987-MCinHF,Ni}.
\begin{figure}
\includegraphics[width = 5.8 cm]{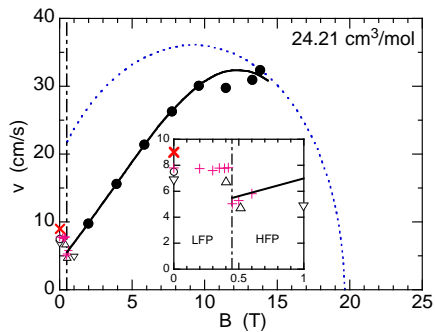}
\caption{Spin wave velocities in the HFP and LFP; 
this work ($\bullet$), Ref.\ \cite{Greywall1987} 
($\triangledown$), Ref.\ \cite{Osheroff1987-MCinHF} ($\triangle$), 
Ref.\ \cite{Ni} (+), Ref.\ \cite{Osheroff-Yu} ($\circ$).
The solid line is a cubic fitting of the present data.
The dotted line is an SWA calculation with Eq.\ (\ref{eq.J}) for the CNAF phase.
The same calculation for the U2D2 phase is also shown ($\times$).
}
\label{fig.SWvelocity}
\end{figure}

Although our linear spin-wave approximation (SWA) calculation for $v$ 
with the MSE parameters in Eq.\ (\ref{eq.J}) (the dotted line in Fig.\ 4) 
agrees with the experiment at the highest fields, 
it gives too large values in lower fields, e.g., four times larger at $B_{c1}$.
In order to obtain such small $v$ values at low fields, quite large $J_{2N}$ and $S_{1}$ 
even comparable to $J_{1N}$ have to be assumed in the calculation, which is unreasonable.
Thus the discrepancy indicates that the spin-wave excitation is ill defined 
in the HFP in low fields due to the frustration and 
the resultant large fluctuations.
The fluctuations are expected to be larger in the CNAF phase than 
in the U2D2 phase because of the non-collinear spin structure 
in the former phase.
Eventually, the same SWA calculation for the U2D2 phase yields 
$v =$ 9.0 cm/s which is in reasonable agreement with the experimental one 
(= 7.7-7.8 cm/s) \cite{Ni,Osheroff-Yu}.
To test this explanation, it is worthwhile to perform the next-order spin 
wave calculations for both the ordered phases.

In conclusion, we studied the nuclear AFM orderings of the melting 
bcc $^3$He in high fields.
The observed reentrant nature of the HFP-PP boundary 
strongly support the MSE hypothesis excluding other scenarios 
related to the ZPV.
The zero-temperature upper critical field has been determined as 
19.7 T reliably from its volume dependence.
The higher order exchanges than six spins can not be neglected to
account for the present high field data.
We also found the considerable softening of the spin waves in the HFP 
at low fields, which is indicative of strong frustration due to the competing MSEs.
A future extension of this work to higher fields near 20 T and 
a new PIMC calculation of $J_P$ beyond six spins at different densities 
are desirable to verify these conclusions. 

We thank D. Ito, T. Morita, T. Fukuda, T. Okamoto and S. Ogawa 
for their contributions to this experiment.
This work was financially supported by Grant-in-Aid for Scientific Research 
from MEXT, Japan.


\begin{thebibliography}{99}
\bibitem{unpublished}
The HFP-PP transition data at 22.47 and 23.06 cm$^3$/mol  
were obtained with the same experimental setup as Ref.\ \cite{Fukuyama-Bc2}.
\end{thebibliography}
\end{document}